\documentclass{article}
\usepackage[T1]{fontenc} 
\usepackage[utf8]{inputenc} 
\usepackage{ismir,amsmath,cite,url}
\usepackage{graphicx}
\usepackage{color}
\usepackage{amsmath}
\usepackage{graphicx}
\usepackage{amssymb}
\usepackage{musicography}
\usepackage{float}
\usepackage{subcaption}
\usepackage{color}
\usepackage{tablefootnote}
\usepackage[algo2e]{algorithm}
\usepackage{algorithmic}
\usepackage{multicol}
\usepackage[dvipsnames]{xcolor}

\title{A New Dataset, Notation Software, and Representation for Computational Schenkerian Analysis}

\multauthor
{
Stephen Ni-Hahn$^1$ \hspace{1cm} 
Weihan Xu$^1$ \hspace{1cm} 
Jerry Yin$^1$ \hspace{1cm} 
Rico Zhu$^1$ 
} { 
\bfseries{ \hspace{1cm} Simon Mak$^1$ \hspace{1cm} Yue Jiang$^1$ \hspace{1cm} Cynthia Rudin$^1$}\\
 $^1$Duke University, USA\\
{\tt\small stephen.hahn@duke.edu}
}

\def\authorname{F. Author, S. Author, and T. Author}

\begin{document}

\maketitle
\begin{abstract}
\textit{Schenkerian Analysis} (SchA) is a uniquely expressive method of music analysis, combining elements of melody, harmony, counterpoint, and form to describe the hierarchical structure supporting a work of music. However, despite its powerful analytical utility and potential to improve music understanding and generation, SchA has rarely been utilized by the computer music community. This is in large part due to the paucity of available high-quality data in a computer-readable format. With a larger corpus of Schenkerian data, it may be possible to infuse machine learning models with a deeper understanding of musical structure, thus leading to more ``human'' results. To encourage further research in Schenkerian analysis and its potential benefits for music informatics and generation, this paper presents three main contributions: 1) a new and growing dataset of SchAs, the largest in human- and computer-readable formats to date (>140 excerpts), 2) a novel software for visualization and collection of SchA data, and 3) a novel, flexible representation of SchA as a heterogeneous-edge graph data structure.

\end{abstract}
\section{Introduction}\label{sec:introduction}

With the continuously growing availability of ``big data,'' machine learning models and algorithms have made enormous strides in many fields, such as computer vision and language modeling. Recent approaches to music information retrieval (MIR) and music generation tasks are increasingly fueled by massive datasets as well, particularly when working with raw audio. For instance, for generation tasks, Meta's \textit{MusicGen} is trained on approximately 20,000 hours of licensed music \cite{copet2024simple}, OpenAI's \textit{Jukebox} on 1.2 million songs \cite{dhariwal2020jukebox}, and Google's \textit{Noise2Music} on 340,000 hours of music \cite{huang2023noise2music}. Castellon et al. show how these large generation models produce useful representations for downstream MIR tasks \cite{castellon2021codified}. Won et al. perform multimodal metric learning for tag-based music retrieval using approximately 158,000 tracks\cite{won2021multimodal}. 

Despite this promising body of work, many areas of music research do not have access to such data and are therefore under-researched and underappreciated, particularly in the realm of symbolic music or Schenkarian Analysis (SchA). By infusing an understanding of Schenkerian musical structure, generative machine learning models may be able to learn more artistic, theoretically-informed structural features beyond simple form and metric features when making inference. Unfortunately, there is currently only one sizeable publicly available dataset for SchA in computer-readable format, and it is relatively small with 41 excerpts\cite{kirlin2014data}.

Schenkerian analysis provides a powerful, flexible, and broadly-used analytical framework for understanding musical melodic-harmonic structure in a sensitive, ``human'' way. Rather than viewing a piece of music as a series of vertical chunks or horizontal melodies, SchA instead analyzes music as an artistic ``unfolding'' of harmony through time, taking into account elements of melody, harmony, form, and counterpoint. Schenker's theories have inspired numerous performers and composers \cite{Finane, schenker2000art, jackson2001heinrich}, helping them to understand their own interpretations of musical structure, which in turn may inform their own performance and/or composition. An understanding of Schenkerian structure helps performers determine what notes deserve emphasis and which may be more transient. A composer can learn to imitate and develop structures they have seen in other pieces of music they admire. 

Because Schenkerian theory requires a significant amount of background knowledge in music theory and practice and has a difficult learning curve, it is often overlooked or misunderstood. For instance, SchA is often deemed too narrow due to its origins in repertoire of Western common practice tonal music. However, aspects of Schenkerian theory have shown strong analytical power in works of popular, rock, jazz, and even African folk music, Chinese opera, and 20th century western atonal music\cite{nobile2014, stock1993application, larson2009analyzing, didier2022form}. To be clear, we see SchA as a broad and evolving field with various analytical tools that can be applied to a wide array of musical genres, not as a static theory solely designed for common practice tonality. 


It is our belief that research in computational SchA can enable performers and composers to more easily analyze music and guide the process of understanding musical structure. Computational SchA can also aid the expert human analyst by offering several potential solutions, speeding up their ability to parse through a piece of music or potentially unveiling unusual and interesting analytical insights. The computer would not replace the human expert; rather, it would help the analyst find reasonable solutions more quickly, which would be immensely helpful when conducting large-scale corpus studies. Furthermore, inclusion of SchA in MIR and generation tasks may greatly improve results. This injection of computational models with musical theory and/or structure has shown benefits in numerous MIR and generation tasks \cite{fong2023theory, wu2019hierarchical, hahn2023interpretable, zhang2022structure}.


This paper introduces three main contributions. The first is a new and growing \textbf{SchA dataset}, the largest in human- and computer-readable format to date (>140 excerpts). Second, we present a novel \textbf{notation software for SchA} in an effort to ease data collection and visualization. Lastly, we describe a \textbf{representation of SchA} as a graphical data structure and graph pooling problem.

The following subsections describe SchA in more detail, as well as the relevant history of computational SchA. Section \ref{sec:data} describes our novel dataset and data collection tool. Finally, Section \ref{sec:methodology} describes how SchA may be represented as a graph data structure.

\subsection{Hierarchical Music Analysis}\label{subsec:hierarchical_analysis}

\begin{figure}[H]
    \centering
    \includegraphics[width=0.40\textwidth]{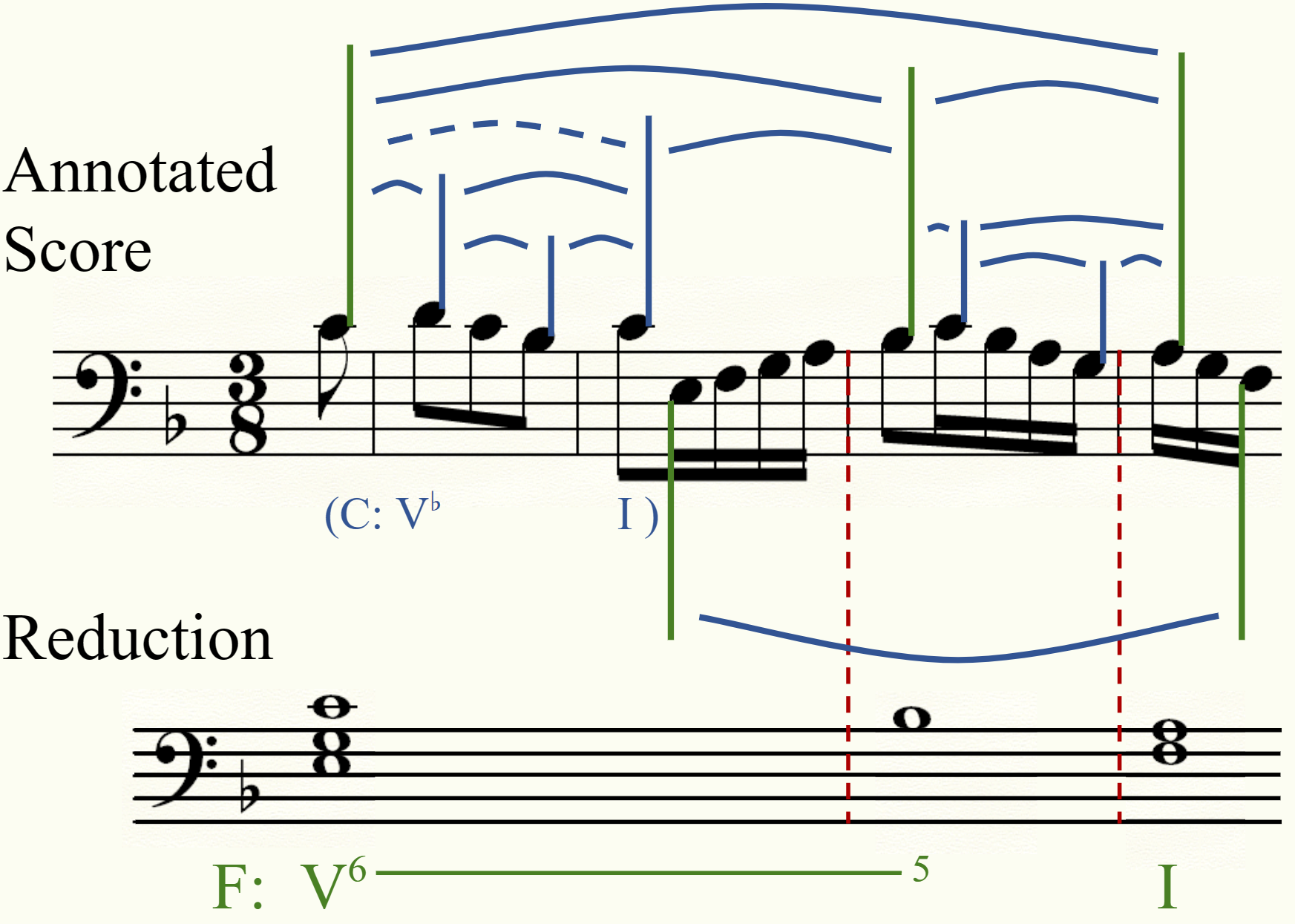}
    \vspace{-3mm}
    \caption{The primary author's analysis of J.S. Bach's F major fugue subject from \textit{Das Wohltemperierte Klavier I}.}
    \label{fig:bach_analysis}
\end{figure}

Music is often composed and understood in terms of hierarchical structures such as phrase and rhythmic structure\cite{rothstein1989phrase, lerdahl1996generative}, form structure \cite{hepokoski2006elements, caplin2013analyzing}, and linear/harmonic structure \cite{schenker2001free, cadwallader1998analysis}. In this paper, we focus on the Schenkerian model of harmonic-melodic structure. As mentioned earlier, SchA aims to reveal how harmonies are ``unfolded'' through linear motion on various levels of structure. Figure \ref{fig:bach_analysis} shows the relationship between a fugue's subject melody and its underlying harmony, as well as the hierarchy of melodic tones.

\noindent The annotated score on the upper line shows how notes relate on various levels of structure, forming two theoretical outer voices. Longer stems indicate deeper levels of structure. The reduction on the bottom line exemplifies the underlying harmony that is unfolded by the subject melody. \textcolor{ForestGreen}{Green}-stemmed notes correspond with the deep outer voices of the reduction.

SchA has shown that similar harmonic and motivic features often exist on multiple levels of hierarchy, revealing music's ``fractal'' nature\cite{burkhart1978schenker}. For instance, in Figure \ref{fig:bach_analysis}, the foreground melody within the first full measure (D4-C4-B$\flat$3) can be seen as a \textit{parallelism} of the deep level melody spanning the entire excerpt (C4-B$\flat$3-A3); the two melodies have a similar motivic descending third in step-wise motion. One can also see the first full measure leading into the second measure as a $\text{V}^{\flat}-\text{I}$ motion in the key of V, paralleling the deep level $\text{V}^\natural-\text{I}$ shown in the reduction. While these examples are on a very small scale, one can see more complex harmonic and motivic structures unfolded through entire pieces. For instance, see Example 12 in \cite{burkhart1978schenker} describing Schubert's \textit{Erlk{\"o}nig} or Example 2 in \cite{gauldin1990beethoven} describing The Beatles' \textit{Something}.

Because these same music-theoretical ideas and motifs permeate multiple levels of structure, the use of a carefully-designed machine learning model may reveal such structure in a layered approach. With the rise of machine learning in data science, this calls into importance the need for computer-readable SchA datasets for model training.

\subsection{Previous Work and Data for Computational Schenkerian Analysis}\label{subsec:computational_schenker}

The majority of past attempts at computational SchA \cite{kassler1975proving, frankel1976lisp, smoliar1979computer, mavromatis2004parsing, gilbert2007probabilistic} were based on heuristics and rule-based algorithms, and therefore did not require a true computer-readable dataset for SchA. To our knowledge, Marsden \cite{marsden2010schenkerian} was the first to venture towards a machine learning approach, using a humble corpus of six Mozart analyses. He developed a ``goodness metric'' based on these six analyses to find the best candidate analyses within a massive search space. 

More recently, Kirlin designed a probabilistic model for SchA that understands SchAs as maximal outerplanar graphs (MOPs) and learns how likely certain notes \textit{prolong} others using random forests\cite{kirlinDissertation, yust2006formal}. He defines prolongation as ``a situation where an analyst determines that a group of notes is elaborating a group of more structurally fundamental notes.'' For instance, the syntax follows the pattern $X (Y) Z$, where the note(s) of $Y$ prolong the motion from note $X$ to note $Z$. 

One potential drawback of Kirlin's model is that it always considers one musical voice as one theoretical voice. Looking back at Figure \ref{fig:bach_analysis}, for example, we see there is clearly a deep level bass motion from E3 to F3, supporting the upper voice, which follows the motion C4-B$\flat$3-A3. The sixteenth notes of m$.$ 2 act to fill the gap between the lower and upper theoretical voices. An MOP reduction of the melody would force all notes to be understood as a single theoretical voice, thus obscuring the underlying counterpoint of the passage. For this reason, we represent SchA as a more general graph data structure, described further in Section \ref{sec:methodology}.

\begin{figure*}[h!]
\centering
\begin{subfigure}{.5\textwidth}
  \centering
  \hspace{-10mm}
  \includegraphics[width=\linewidth]{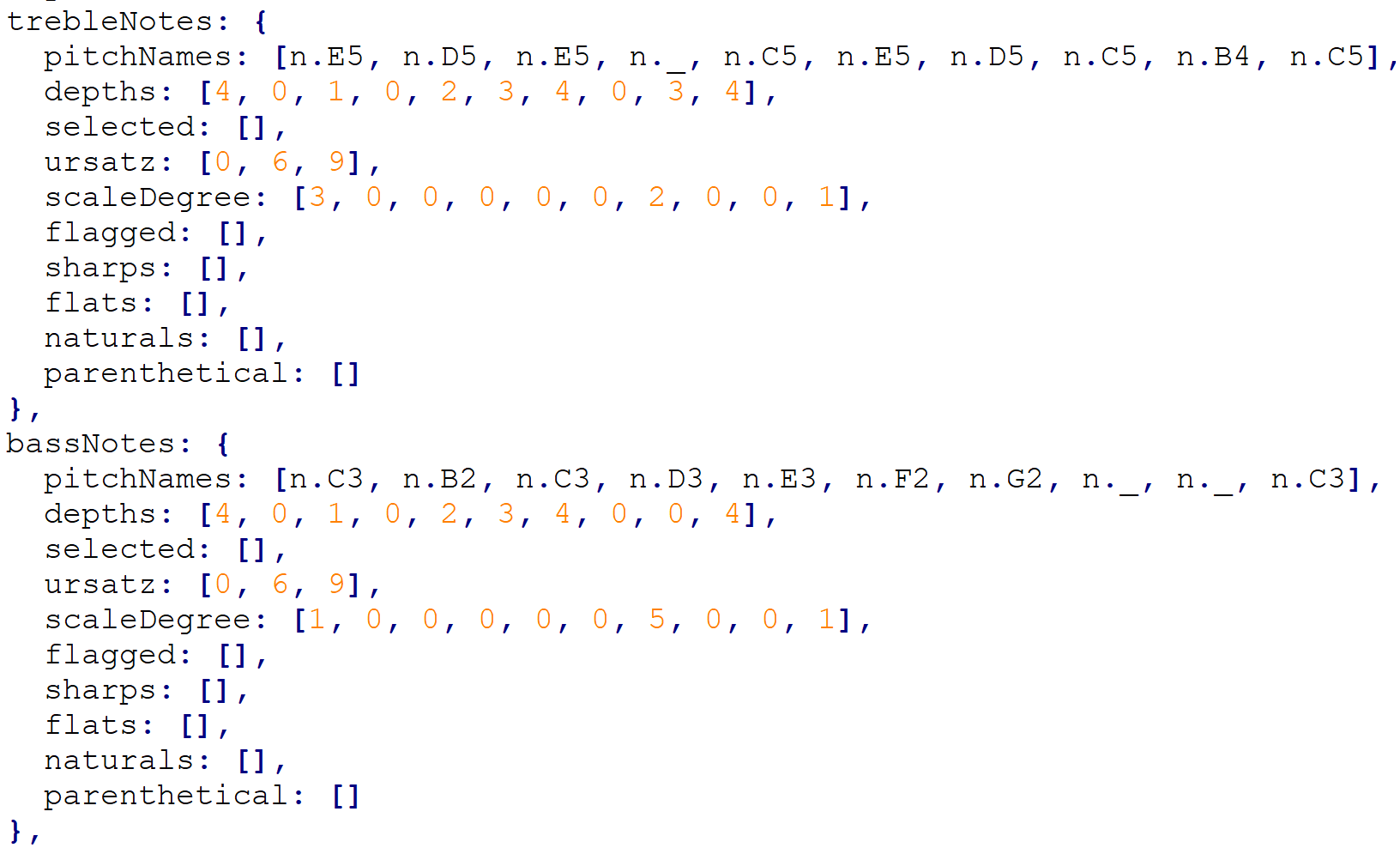}
  \caption{JSON representation.}
  \label{fig:sub1}
\end{subfigure}%
\begin{subfigure}{.5\textwidth}
  \centering
  \includegraphics[width=\linewidth]{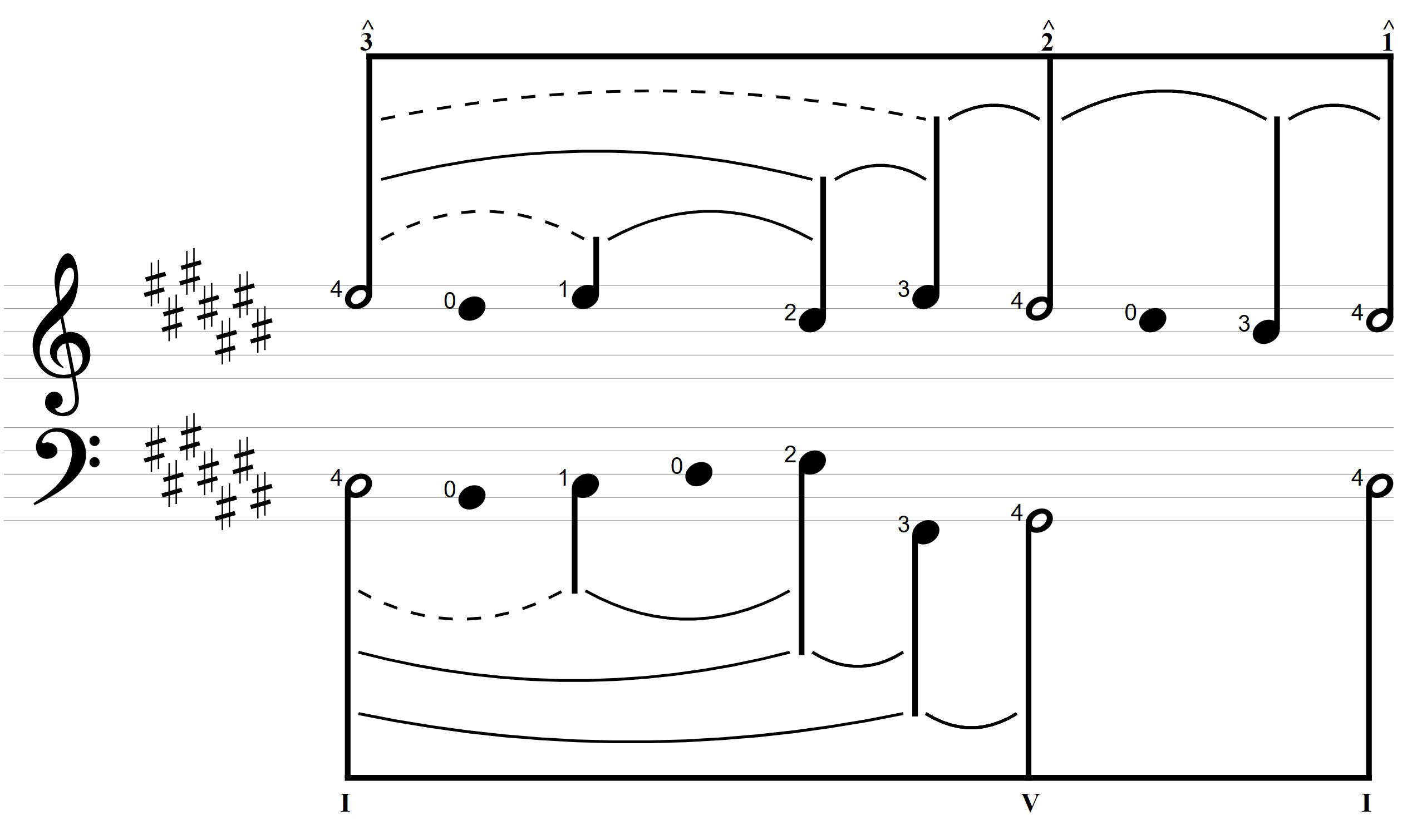}
  \caption{Graphical representation.}
  \label{fig:sub2}
\end{subfigure}
\caption{Screenshots of a toy Schenkerian analysis in JSON and graphical form as generated by our notation software.}
\label{fig:notation_tool}
\end{figure*}

For his model, Kirlin released the first large-scale computer-readable dataset of symbolic music with corresponding expert Schenkerian analyses, \textit{Schenker41} \cite{kirlin2014data}. This collection consists of 41 excerpts from the common practice period of European art music, with analyses from three textbooks \cite{forte1982instructor, forte1982introduction, cadwallader1998analysis, pankhurst2008schenkerguide} and an independent, anonymous expert in Schenkerian analysis. Kirlin also created the first computer-readable format for Schenkerian analysis, which describes all prolongations present in an analysis. The text-based format can also encode linear progressions, omitted repetitions, and harmonic context.

The Schenker41 dataset is an important first step towards broader musical-hierarchical research in the MIR community; however, there are some limitations. First of all, the quality of the excerpts chosen are questionable. Kirlin and Jensen recruited three expert Schenkerian analysts to grade textbook analyses as well as their machine learning model's analyses in their 2015 paper (see Figure 8 in \cite{kirlin2015learning}). One would expect the textbook analyses to receive grades of ``A-'' or greater, allowing wiggle room for differences of opinion. However, many excerpts score lower marks; some were even given failing grades. Given the high proportion of dubious quality ``ground truth'' data, it is necessary to produce a greater quantity of quality data before successful, generalizable models can be trained.

There are also several Schenkerian symbols and concepts that are not currently represented in the text-based notation. For instance, unfoldings, voice exchanges, and other hierarchical harmonic function information are ignored. Concerning larger pieces, it is vital to understand the harmonic structure in several layers; an F major triad may stand as a local tonic ``I'' harmony in the foreground that serves to expand a deeper subdominant ``IV'' of the background, global key of C. Furthermore, more abstract concepts, such as motivic parallelisms, implied tones, and written commentary are eschewed for the sake of simplicity.

\section{Dataset and Notation Software for Schenkerian Analysis}\label{sec:data}

We thus introduce a new large-scale dataset of Schenkerian analyses in human- and computer-readable formats. As of the writing of this paper, the dataset contains 145 analyses from four analysts for a broad range of composers including J.S. Bach, Mendelssohn, Brahms, Bart\'{o}k, Shostakovich, Gentle Giant, and more. The majority of analyses were created by the first author (Stephen Ni-Hahn) with consultation from one of the other analysts, who wishes to remain anonymous at this time. Ni-Hahn currently has nearly a decade of experience with SchA including a graduate degree in music theory. The other three analysts are veteran Schenkerian scholars with several decades of experience in the field. The dataset is not static and aims to grow over time. Please contact Stephen Ni-Hahn (stephen.hahn@duke.edu) for questions regarding, and access to the dataset and notation software described in this paper.

Currently, the vast majority of analyses in the dataset describe the hierarchical relationships within fugue subjects by Bach and Pachelbel. Fugue subjects are ideal for preliminary trials with machine learning models since subjects are generally brief, consist of a single instrumental line (which may consist of multiple theoretical voices), generally have clear functional relationships, and each have a definite sense of closure by their end.

\begin{figure*}[h!]
\centering
\begin{subfigure}{.4\textwidth}
  \centering
  \hspace{-4mm}
  \includegraphics[width=\linewidth]{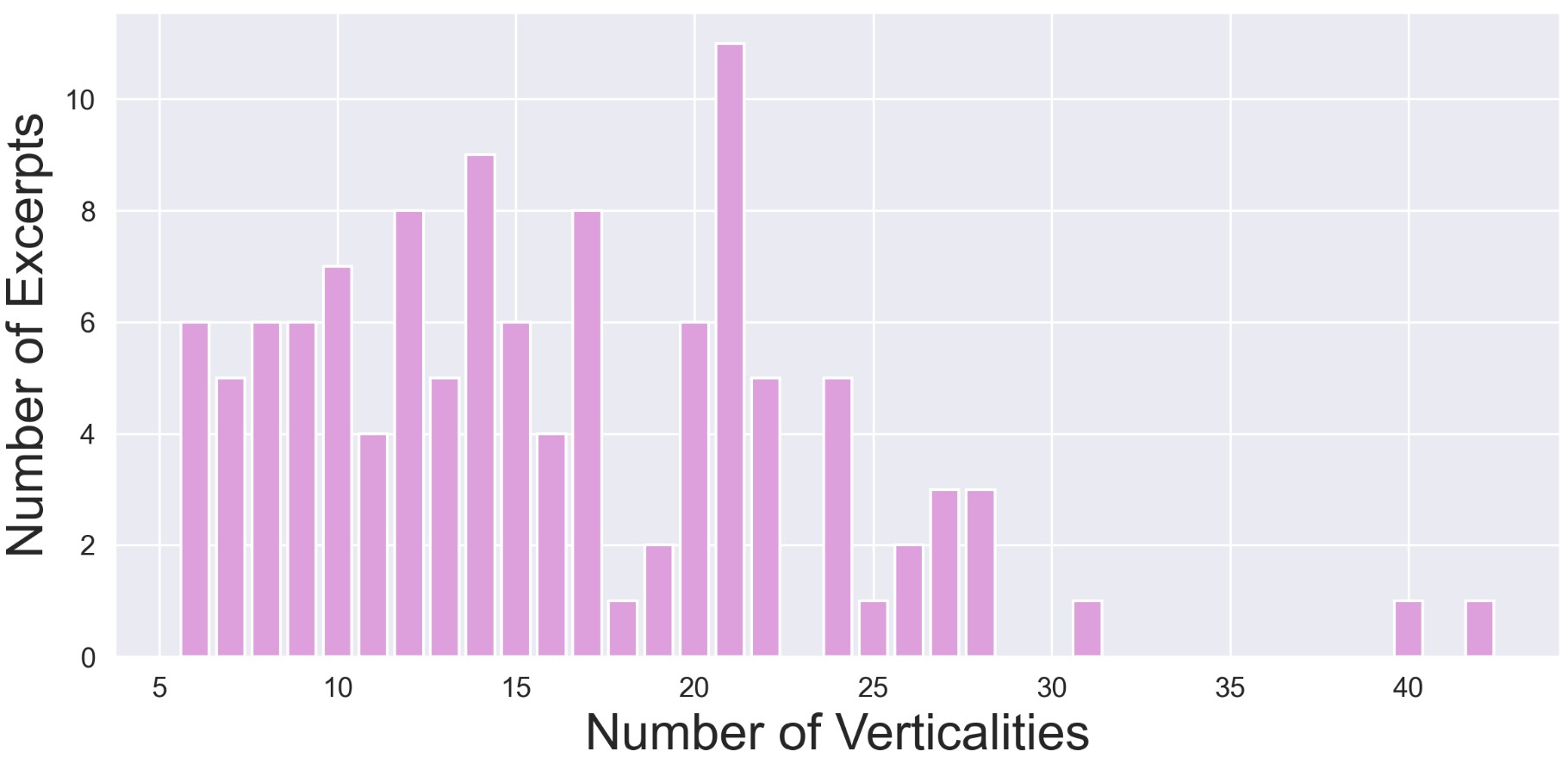}
  \caption{Distribution of excerpt length. }
  \label{fig:verticalities}
\end{subfigure}%
\begin{subfigure}{.6\textwidth}
  \centering
  \includegraphics[width=\linewidth]{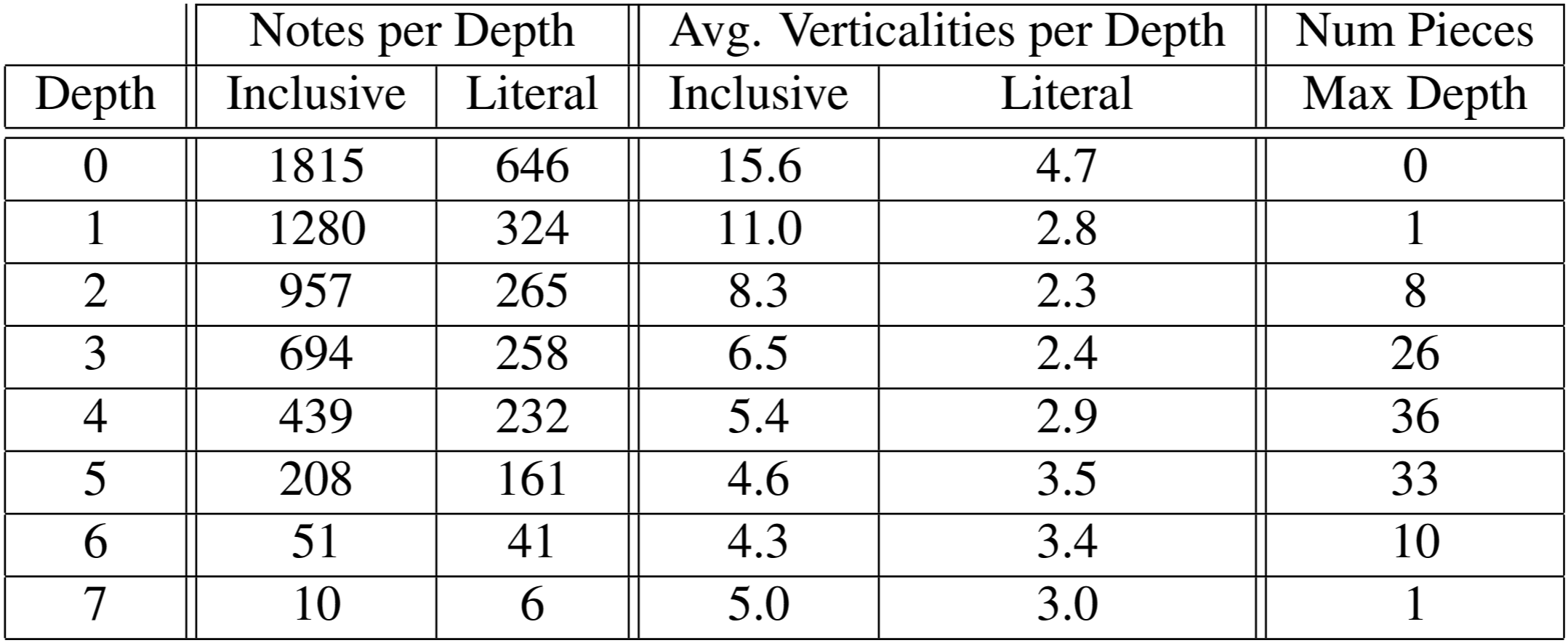}
  \caption{Dataset statistics regarding note depth.}
  \label{table:stats}
\end{subfigure}
\caption{Dataset statistics. Verticality is defined as a point in time where one or both of a treble and bass note exist. ``Inclusive'' includes notes of higher depth when counting notes of lower depths. ``Literal'' counts the note depths as they are defined. The final column describes the distribution of max depths over all excerpts. See Section \ref{sec:data} for more details.}
\label{fig:stats}
\end{figure*}

Rather than writing out each prolongation explicitly, we produce prolongations as a by-product when assigning a hierarchical \textit{depth} to each note. For example, Figure \ref{fig:notation_tool} shows a toy example of an analysis in which the numbers to the left of the note heads indicate depth. Higher depth indicates deeper structure. To retrieve the prolongations, we simply traverse the graph at each depth level (greater than 0), connecting consecutive notes that are at the same level or higher. Custom  prolongations that do not occur within this system may be added in a similar fashion to Kirlin's text format by describing the voice and index of the start, middle, and end notes.

\begin{figure}[H]
    \centering
    \includegraphics[width=0.48\textwidth]{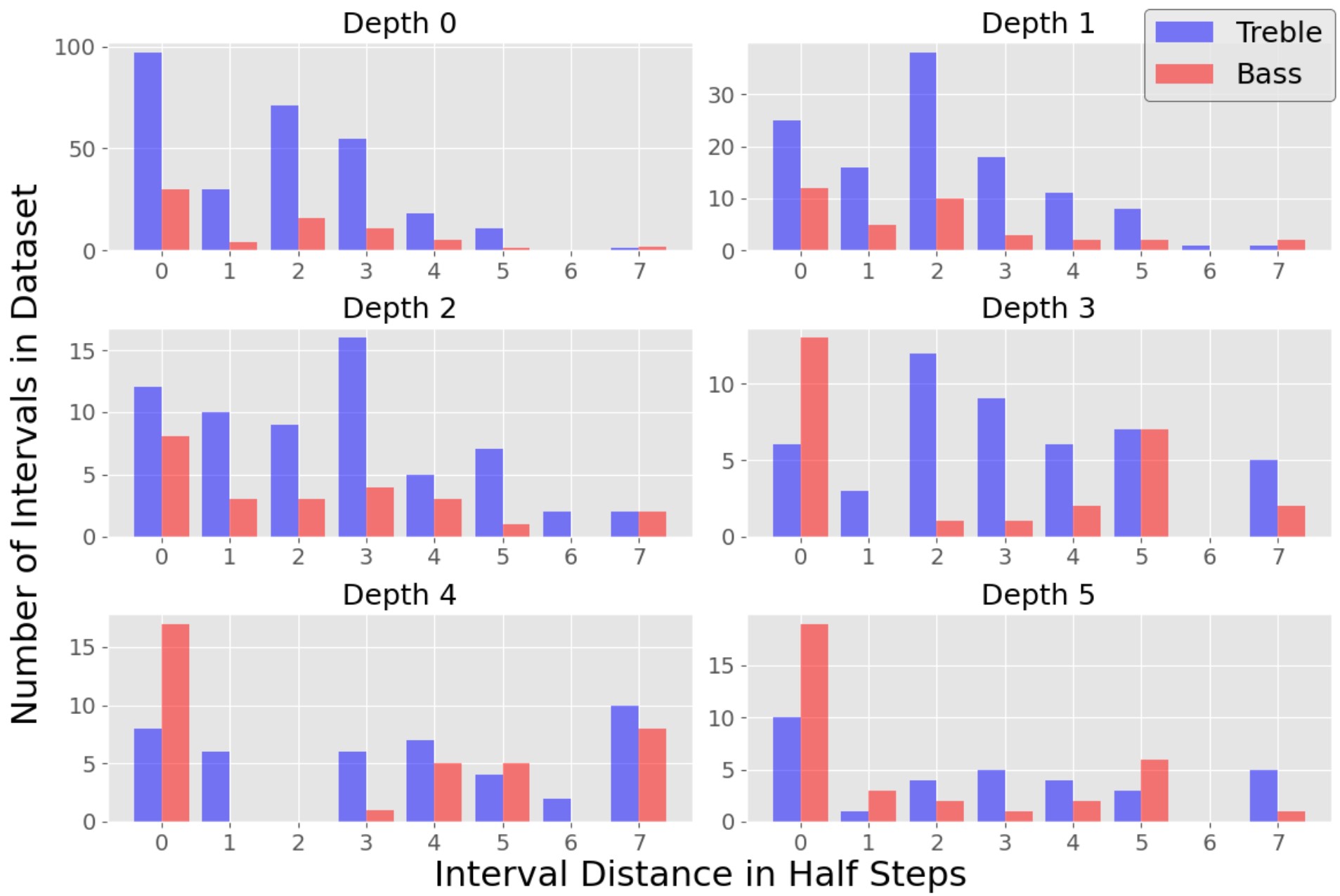}
    \caption{Distribution of intervals between consecutive notes at each depth.}
    \label{fig:intervals}
\end{figure}

Figures \ref{fig:stats} and \ref{fig:intervals} present simple statistics about our dataset. Figure \ref{fig:verticalities} shows the distribution of excerpt lengths in terms of \textit{verticalities}. A verticality is defined as a point in time where one or both of a treble and bass note exist. Note that this does not measure length of time or number of measures; rather, the number of verticalities describes the number of potentially unique depths in an excerpt. Figure \ref{fig:intervals} shows the distribution of intervals between consecutive notes in the treble and bass voices at various depths. We see that as depth increases, the distribution of treble intervals moves from smaller to larger intervals, while bass intervals increasingly concentrate around 0 and 5. These statistics suggest that surface level treble motions in our dataset are mostly stepwise and span larger intervals at deeper levels of structure. Furthermore, deep bass structures tend to hold steady and support the upper voice or move along the circle of fifths by jumping 5 or 7 half steps. Table \ref{table:stats} describes various statistics regarding the notes and depths of our dataset. Columns labeled ``inclusive'' mean that notes of higher depth are included when counting notes of lower depths. For instance, a depth 4 note is counted in the number of depth 0 notes, while the depth 4 note would not count towards the number of depth 5 notes. The ``literal'' label counts the note depths as they are defined. The final column describes the distribution of max depths over all excerpts. 



\subsection{Data Collection Tool}\label{subsec:data_collection_tool}

To facilitate easy collection and visualization of Schenkerian data, we introduce a new computer notation system for Schenkerian analyses (see Figure \ref{fig:notation_tool} for a screenshot). 

As of the writing of this paper, our software is capable of notating up to four voice structures of any length. Simple commands allow the user to adjust the pitches, note depths, harmonic/scale-degree label, notes considered part of the \textit{Ursatz}, etc. Slurs and beams of the outer theoretical voices are automatically generated based on the depths of the notes. We are currently working on ways to render custom markings, such as voice exchanges, unfoldings, and linear progression beams.

Behind the scenes, the Schenkerian analysis is a simple standardized object in JavaScript Object Notation (JSON), which is \textit{highly generalizable, lightweight, and simple to parse}, and is capable of describing any obscurities within a particular analysis. Our JSON object contains metadata about the analysis, key information, and information on each of four theoretical voices. Metadata describes the analyst, composer, title, subtitle, and any associated written description of the analysis. Furthermore, each theoretical voice is encoded as a list of pitch names, depths, \textit{Ursatz} indices, scale degree/Roman numerals, flagged note indices, sharp/flat/natural indices, and parenthetical indices. Additionally, the JSON object stores ``cross voice'' symbols such as voice exchange lines and lines indicating related tones across larger spans of time. 

Note that it is straightforward to translate between Kirlin's OPC text notation and our JSON notation. To translate from text to JSON, the notes can be parsed from the musicxml and placed in their appropriate voice. Then note depths may be determined by the location and relative length of their prolongation. Translating from JSON to text is simpler, as one can traverse each depth and retrieve the prolongations.

The software is constructed using languages Javascript/Typescript and the Vue web framework. It is packaged using Electron Forge. Software access can be requested by emailing the first author.

\section{Schenkerian Analysis as a Heterogeneous Graph Data Structure}\label{sec:methodology}

\begin{algorithm*}
\caption{JSON to Clusters }\label{algorithm:json_to_clusters}

\textbf{Definitions}

\begin{algorithmic}[h]
    \STATE $parts \leftarrow \{sop, \;alto, \;ten, \;bass\}$
    \STATE $n_v \leftarrow$ the number of verticalities $v$ (indexed by $i$) in an analysis
    \STATE $p_i \leftarrow$ note of part $p \in parts$ within $v_i$
    \STATE $d_i^{(p)} \leftarrow$ depth of note $p_i$
    \STATE $\forall p \in parts$, $\text{len}(p) = \text{len}(d^{(p)}) = n_v$.

\end{algorithmic}

\bigskip

\textbf{Procedure} CLUSTER($p,i$)
\begin{algorithmic}
    \IF{$\exists j < i \, \text{s.t.} \, d_j^{p} > 0$}
        \STATE $j\leftarrow\underset{j}{\mathrm{argmin}} \, |i - j| \quad \text{s.t.} \quad j<i \;\;\text{and}\;\; d_j^{(p)} > 0$
        \RETURN $\{(p,j)\}$ $\hfill \text{// Note in the same voice to the left}$
    \ELSIF{$\exists j > i \, \text{s.t.} \, d_j^{p} > 0$}
        \STATE $j\leftarrow\underset{j}{\mathrm{argmin}} \, |i - j| \quad \text{s.t.} \quad j>i \;\;\text{and}\;\; d_j^{(p)} > 0$
        \RETURN $\{(p,j)\}$ $\hfill \text{// Note in the same voice to the right}$
    \ELSE
        \STATE $j_1,j_2\leftarrow\underset{j_1,j_2}{\mathrm{argmin}} \, min(|i-j_1|,|i-j_2|)\quad \text{s.t.} \quad (i-j_1)\cdot (i-j_2)\le 0 \;\;\text{and}\;\; d_{j_1}^{(sop)}>0 \;\;\text{and}\;\; d_{j_2}^{(bass)}>0$
        \RETURN $\{(sop,j_1),(bass,j_2)\}$ $\hfill \text{// Closest two notes in outer voices in opposite directions to the inner voice note}$
    \ENDIF
\end{algorithmic}
\end{algorithm*}

As mentioned in Section \ref{subsec:computational_schenker}, Kirlin's model simplifies the difficult problem of performing SchA, using a limited version of Yust's MOP representation for SchA. With a greater amount of data, less compromising representations may be used for modeling. The following section describes how a musical score may be represented as a heterogeneous-edge directed graph data structure and how SchA may be conceptualized as a graph clustering problem.

\subsection{Graph Music Representation}\label{subsec:music_representation}

\begin{figure*}[h!]
    \centering
    \hspace*{-5mm}  
    \includegraphics[width=\textwidth]{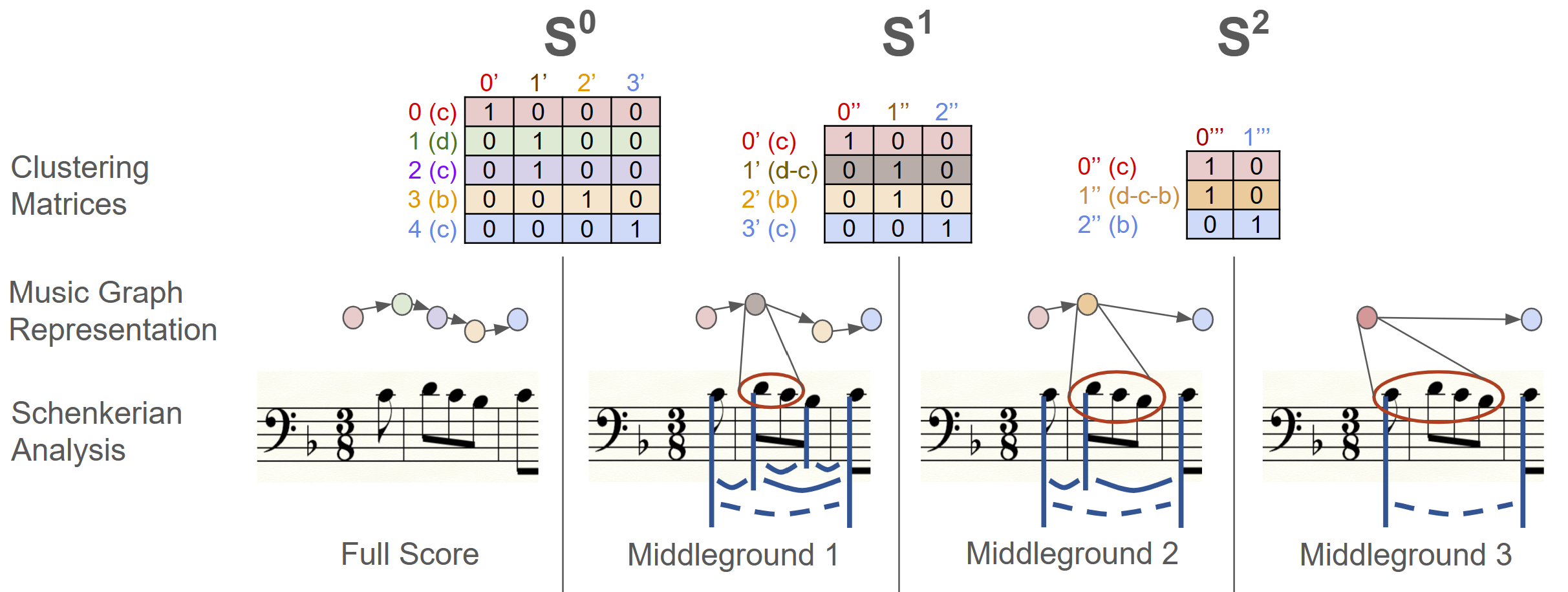}
    \vspace{-3mm}
    \caption{Visualization of Schenkerian analysis as a series of clustering matrices. The bottom row shows a simple score with Schenkerian annotation moving from all notes in the score to more abstracted versions of the score from left to right. The middle row visualizes the music as a graph. The top row shows the ground truth cluster matrices that relate one layer to the next; rows describe nodes before clustering, while columns describe nodes after clustering.}
    \label{fig:clustering_visualization}
\end{figure*}

In what follows, we represent music as a heterogeneous directed graph $G$, where each node describes a note, and various types of edges describe the relationships between notes. Concretely, $G$ is represented as ($\mathbb{A}$, $X$), where $\mathbb{A} \in \{0, 1\}^{h\times n \times n}$ describes the set of $h$ adjacency matrices (one for each edge type) over $n$ nodes, and $X \in \mathbb{R}^{n \times d}$ is the node feature matrix with $d$ as the number of features. These $d$ features may be learned by a neural network, for instance, to correspond with categorical and numerical musical features.

We adapt the encoding scheme proposed by Jeong et al.\cite{jeong2019graph} for the purpose of Schenkerian analysis. Nodes may be encoded with any musical feature present in the score, such as pitch class, octave, absolute duration, position (absolute or relative), metric strength, etc. We suggest the use of five main edge types: (i) forward edges connect two consecutive notes within a voice, (ii) onset edges connect notes that begin at the same time, (iii) sustain edges connect notes that are played while the source note is held, (iv) rest edges are like forward edges, but imply a rest occurs between the two related notes, and (v) linear edges connect each note with the next notes that occur at specific intervals from the source. 


\subsection{Schenkerian Analysis as Hierarchical Clustering}\label{subsec:SchA_as_clustering}

With this graphical representation of music, the process of Schenkerian analysis may then be posed as a hierarchical graph clustering problem. Figure \ref{fig:clustering_visualization} presents a toy example of how Schenkerian analysis may be represented as a series of hierarchical clusters. The clustering between two subsequent levels of Schenkerian analysis is expressed through a \textit{clustering matrix}, $S^{(l)} \in \mathbb{R}^{n_l \times n_{l+1}}$, where $n_l$ is the number of nodes in \textit{clustering} layer $l$ and $n_{l+1} < n_l$ is the number of nodes after one iteration of clustering. We define $n_0$ to be the total number of notes in the music.

Note that we can understand a clustering between \textit{any} two layers as a single matrix, denoted as $S^{(l_i)\rightarrow(l_j)} \in \mathbb{R}^{n_{l_i} \times n_{l_j}};\; i<j$, where $i$ and $j$ are the index of the source and destination layers respectively. This single matrix is obtained by simply multiplying all sequential clustering matrices. For example, in Figure \ref{fig:clustering_visualization}, to retrieve the matrix describing how all five nodes of the original score are clustered into the two nodes of the final middleground layer, we can multiply each clustering matrix together:

\vspace{-2mm}

$$S^{(0)\rightarrow(2)} = S^{(0)}\cdot S^{(1)} \cdot S^{(2)} = 
\begin{bmatrix}
1 & 1 & 1 & 1 & 0\\
0 & 0 & 0 & 0 & 1\\
\end{bmatrix}^\top
.$$

\subsection{Converting Schenkerian Analyses from JSON to Matrix Notation}\label{subsec:json_to_matrix}

Schenkerian analysis JSON data (collected using our tool described in Section \ref{subsec:data_collection_tool}) requires extra processing to be represented as hierarchical clusters. Here, we provide an algorithm to convert our JSON data into a series of progressively smaller clustering matrices (see Algorithm \ref{algorithm:json_to_clusters}). 

Essentially, we first traverse the outer voices of the JSON file, clustering notes of depth 0 into the closest note of higher depth to the left in the same voice. If that note does not exist, it defaults to the closest note of a higher depth to the right. For inner voices, if they do not describe hierarchical depth (all 0 depth), they are clustered 50\%-50\% between the nearest bass and soprano below and above or left to right, in that order. If the inner voice has specified depth, it is treated similarly to the outer voices. All depths are then decremented and the process begins again for the next clustering matrix.

\subsection{Implications of SchA Graph Clustering}\label{subsec:pros_cons}

The above formulation of SchA as a graph clustering problem facilitates more generalizable analysis. Whereas Kirlin's MOP-based model focuses on a single melody as one theoretical voice, a fuller graph representation allows for greater flexibility via any number of theoretical voices. There are, however, several drawbacks with this new approach. Because the clustering works with the notes of the score, it is unclear how to handle cases where multiple theoretical voices converge on a single note. This issue may also be present when handling inner voices of unspecified depth. In our algorithm, we suggest splitting unspecified inner voices 50\%-50\% between the outer voices, but other approaches may also be reasonable. 

Another advantage that the proposed graph clustering representation has over the MOP representation is its ability to cluster multiple notes into one in a single layer. This is particularly common when there are several repeated notes. In a MOP, repeated notes must be given detailed hierarchy, whereas a human expert would generally think of such repetitions as structurally redundant. There are also instances of prolongations that span more than one child, where having only one child would not properly reflect the music. For instance, if the melody over a C major tonic triad (CEG) quickly plays out the upper tetrachord of the scale, G-A-B-C, then the A and B are structurally equal; they both bridge the gap from G to C. On the other hand, allowing multiple children for every prolongation makes the search space for potential solutions orders of magnitude larger.

As the amount of labeled SchA data grows and computational power improves, there is great potential for learning complex relationships via machine learning that may be unattainable in previous analyses. Deep learning has enjoyed considerable success on analyzing the Bach chorale dataset \cite{hadjeres2017deepbach, huang2018improved, liang2017automatic}, thus we are optimistic that SchA can also be learned for broad datasets from different genres. The proposed dataset, notation software and graph representation provides a promising step towards this goal.

\section{Conclusion}\label{sec:conclusion}

In this paper, we introduce the largest corpus of Schenkerian analyses in computer-readable format to date. This was largely made possible using our novel SchA notation software, which is natural, interpretable, and enables easy data collection and visualization. Finally, we describe and discuss a novel representation for SchA as a graph clustering problem that allows representation of any possible Schenkerian analysis, avoiding the limitations of MOPs. It is our hope that the growing amount of data and ease of its collection will enable broader research into SchA's applications. 



\bibliography{bibliography}

\begin{thebibliography}{10}
\providecommand{\url}[1]{#1}
\csname url@samestyle\endcsname
\providecommand{\newblock}{\relax}
\providecommand{\bibinfo}[2]{#2}
\providecommand{\BIBentrySTDinterwordspacing}{\spaceskip=0pt\relax}
\providecommand{\BIBentryALTinterwordstretchfactor}{4}
\providecommand{\BIBentryALTinterwordspacing}{\spaceskip=\fontdimen2\font plus
\BIBentryALTinterwordstretchfactor\fontdimen3\font minus \fontdimen4\font\relax}
\providecommand{\BIBforeignlanguage}[2]{{%
\expandafter\ifx\csname l@#1\endcsname\relax
\typeout{** WARNING: IEEEtran.bst: No hyphenation pattern has been}%
\typeout{** loaded for the language `#1'. Using the pattern for}%
\typeout{** the default language instead.}%
\else
\language=\csname l@#1\endcsname
\fi
#2}}
\providecommand{\BIBdecl}{\relax}
\BIBdecl

\bibitem{copet2024simple}
J.~Copet, F.~Kreuk, I.~Gat, T.~Remez, D.~Kant, G.~Synnaeve, Y.~Adi, and A.~D{\'e}fossez, ``Simple and controllable music generation,'' \emph{Advances in Neural Information Processing Systems}, vol.~36, 2024.

\bibitem{dhariwal2020jukebox}
P.~Dhariwal, H.~Jun, C.~Payne, J.~W. Kim, A.~Radford, and I.~Sutskever, ``Jukebox: A generative model for music,'' \emph{arXiv preprint arXiv:2005.00341}, 2020.

\bibitem{huang2023noise2music}
Q.~Huang, D.~S. Park, T.~Wang, T.~I. Denk, A.~Ly, N.~Chen, Z.~Zhang, Z.~Zhang, J.~Yu, C.~Frank \emph{et~al.}, ``Noise2music: Text-conditioned music generation with diffusion models,'' \emph{arXiv preprint arXiv:2302.03917}, 2023.

\bibitem{castellon2021codified}
R.~Castellon, C.~Donahue, and P.~Liang, ``Codified audio language modeling learns useful representations for music information retrieval,'' \emph{arXiv preprint arXiv:2107.05677}, 2021.

\bibitem{won2021multimodal}
M.~Won, S.~Oramas, O.~Nieto, F.~Gouyon, and X.~Serra, ``Multimodal metric learning for tag-based music retrieval,'' in \emph{ICASSP 2021-2021 IEEE International Conference on Acoustics, Speech and Signal Processing (ICASSP)}.\hskip 1em plus 0.5em minus 0.4em\relax IEEE, 2021, pp. 591--595.

\bibitem{kirlin2014data}
P.~B. Kirlin, ``A data set for computational studies of schenkerian analysis.'' in \emph{ISMIR}, 2014, pp. 213--218.

\bibitem{Finane}
\BIBentryALTinterwordspacing
B.~Finane, ``The humanist - murray perahia - steinway \& sons.'' [Online]. Available: \url{https://www.steinway.com/news/features/the-humanist-murray-perahia}
\BIBentrySTDinterwordspacing

\bibitem{schenker2000art}
H.~Schenker, \emph{The art of performance}.\hskip 1em plus 0.5em minus 0.4em\relax Oxford University Press, 2000.

\bibitem{jackson2001heinrich}
T.~L. Jackson, ``Heinrich schenker as composition teacher: The schenker-oppel exchange,'' \emph{Music Analysis}, vol.~20, no.~1, pp. 1--115, 2001.

\bibitem{nobile2014}
\BIBentryALTinterwordspacing
D.~F. Nobile, ``\BIBforeignlanguage{English}{A structural approach to the analysis of rock music},'' Ph.D. dissertation, 2014, copyright - Database copyright ProQuest LLC; ProQuest does not claim copyright in the individual underlying works; Last updated - 2023-03-03. [Online]. Available: \url{https://login.proxy.lib.duke.edu/login?url=https://www.proquest.com/dissertations-theses/structural-approach-analysis-rock-music/docview/1506971540/se-2}
\BIBentrySTDinterwordspacing

\bibitem{stock1993application}
J.~Stock, ``The application of schenkerian analysis to ethnomusicology: problems and possibilities,'' \emph{Music Analysis}, vol.~12, no.~2, pp. 215--240, 1993.

\bibitem{larson2009analyzing}
\BIBentryALTinterwordspacing
S.~Larson, \emph{Analyzing Jazz: A Schenkerian Approach}, ser. ACLS Humanities E-Book.\hskip 1em plus 0.5em minus 0.4em\relax Pendragon, 2009. [Online]. Available: \url{https://books.google.com/books?id=CmMJAQAAMAAJ}
\BIBentrySTDinterwordspacing

\bibitem{didier2022form}
A.~Didier, ``Form and tonal spectrum in 12-tone music: Approaches to analysis in schoenberg, walker, and webern,'' Ph.D. dissertation, University of Oregon, 2022.

\bibitem{fong2023theory}
H.~Fong, V.~Kumar, and K.~Sudhir, ``A theory-based interpretable deep learning architecture for music emotion,'' \emph{Available at SSRN 4025386}, 2023.

\bibitem{wu2019hierarchical}
J.~Wu, C.~Hu, Y.~Wang, X.~Hu, and J.~Zhu, ``A hierarchical recurrent neural network for symbolic melody generation,'' \emph{IEEE transactions on cybernetics}, vol.~50, no.~6, pp. 2749--2757, 2019.

\bibitem{hahn2023interpretable}
S.~Hahn, R.~Zhu, S.~Mak, C.~Rudin, and Y.~Jiang, ``An interpretable, flexible, and interactive probabilistic framework for melody generation,'' in \emph{Proceedings of the 29th ACM SIGKDD Conference on Knowledge Discovery and Data Mining}, 2023, pp. 4089--4099.

\bibitem{zhang2022structure}
X.~Zhang, J.~Zhang, Y.~Qiu, L.~Wang, and J.~Zhou, ``Structure-enhanced pop music generation via harmony-aware learning,'' in \emph{Proceedings of the 30th ACM International Conference on Multimedia}, 2022, pp. 1204--1213.

\bibitem{rothstein1989phrase}
W.~N. Rothstein, ``Phrase rhythm in tonal music,'' \emph{(No Title)}, 1989.

\bibitem{lerdahl1996generative}
F.~Lerdahl and R.~S. Jackendoff, \emph{A Generative Theory of Tonal Music, reissue, with a new preface}.\hskip 1em plus 0.5em minus 0.4em\relax MIT press, 1996.

\bibitem{hepokoski2006elements}
J.~Hepokoski and W.~Darcy, \emph{Elements of sonata theory: Norms, types, and deformations in the late-eighteenth-century sonata}.\hskip 1em plus 0.5em minus 0.4em\relax Oxford University Press, 2006.

\bibitem{caplin2013analyzing}
W.~E. Caplin, \emph{Analyzing classical form: An approach for the classroom}.\hskip 1em plus 0.5em minus 0.4em\relax Oxford University Press, USA, 2013.

\bibitem{schenker2001free}
H.~Schenker, \emph{Free Composition: Volume III of new musical theories and fantasies}.\hskip 1em plus 0.5em minus 0.4em\relax Pendragon Press, 2001, vol.~1.

\bibitem{cadwallader1998analysis}
A.~C. Cadwallader, D.~Gagn{\'e}, and F.~Samarotto, ``Analysis of tonal music: a schenkerian approach,'' \emph{(No Title)}, 1998.

\bibitem{burkhart1978schenker}
C.~Burkhart, ``Schenker's" motivic parallelisms",'' \emph{Journal of Music Theory}, vol.~22, no.~2, pp. 145--175, 1978.

\bibitem{gauldin1990beethoven}
R.~Gauldin, ``Beethoven, tristan, and the beatles,'' in \emph{College Music Symposium}, vol.~30, no.~1.\hskip 1em plus 0.5em minus 0.4em\relax JSTOR, 1990, pp. 142--152.

\bibitem{kassler1975proving}
M.~Kassler, \emph{Proving musical theorems I: The middleground of Heinrich Schenker's theory of tonality}.\hskip 1em plus 0.5em minus 0.4em\relax Basser Department of Computer Science, School of Physics, University of Sydney, 1975, no. 103.

\bibitem{frankel1976lisp}
R.~E. Frankel, S.~J. Rosenschein, and S.~W. Smoliar, ``A lisp-based system for the study of schenkerian analysis,'' \emph{Computers and the Humanities}, pp. 21--32, 1976.

\bibitem{smoliar1979computer}
S.~W. Smoliar, ``A computer aid for schenkerian analysis,'' in \emph{Proceedings of the 1979 annual conference}, 1979, pp. 110--115.

\bibitem{mavromatis2004parsing}
P.~Mavromatis and M.~Brown, ``Parsing context-free grammars for music: A computational model of schenkerian analysis,'' in \emph{Proceedings of the 8th International Conference on Music Perception \& Cognition}, 2004, pp. 414--415.

\bibitem{gilbert2007probabilistic}
{\'E}.~Gilbert and D.~Conklin, ``A probabilistic context-free grammar for melodic reduction,'' in \emph{Proceedings of the International Workshop on Artificial Intelligence and Music, 20th International Joint Conference on Artificial Intelligence}.\hskip 1em plus 0.5em minus 0.4em\relax Citeseer, 2007, pp. 83--94.

\bibitem{marsden2010schenkerian}
A.~Marsden, ``Schenkerian analysis by computer: A proof of concept,'' \emph{Journal of New Music Research}, vol.~39, no.~3, pp. 269--289, 2010.

\bibitem{kirlinDissertation}
P.~B. Kirlin, ``A probabilistic model of hierarchical music analysis,'' PhD thesis, University of Massachusetts Amherst, Amherst, MA, February 2014, available at \url{https://www.cs.rhodes.edu/~kirlinp/diss.html}.

\bibitem{yust2006formal}
J.~D. Yust, \emph{Formal models of prolongation}.\hskip 1em plus 0.5em minus 0.4em\relax ProQuest, 2006.

\bibitem{forte1982instructor}
A.~Forte and S.~E. Gilbert, ``Instructor's manual for introduction to schenkerian analysis,'' \emph{(No Title)}, 1982.

\bibitem{forte1982introduction}
A.~Forte and S.~Gilbert, ``Introduction to schenkerian analysis: structor's manual,'' 1982.

\bibitem{pankhurst2008schenkerguide}
T.~Pankhurst, \emph{SchenkerGUIDE: a brief handbook and website for Schenkerian analysis}.\hskip 1em plus 0.5em minus 0.4em\relax Routledge, 2008.

\bibitem{kirlin2015learning}
P.~Kirlin and D.~Jensen, ``Learning to uncover deep musical structure,'' in \emph{Proceedings of the AAAI Conference on Artificial Intelligence}, vol.~29, no.~1, 2015.

\bibitem{jeong2019graph}
D.~Jeong, T.~Kwon, Y.~Kim, and J.~Nam, ``Graph neural network for music score data and modeling expressive piano performance,'' in \emph{International conference on machine learning}.\hskip 1em plus 0.5em minus 0.4em\relax PMLR, 2019, pp. 3060--3070.

\bibitem{hadjeres2017deepbach}
G.~Hadjeres, F.~Pachet, and F.~Nielsen, ``Deepbach: a steerable model for bach chorales generation,'' in \emph{International conference on machine learning}.\hskip 1em plus 0.5em minus 0.4em\relax PMLR, 2017, pp. 1362--1371.

\bibitem{huang2018improved}
C.-Z.~A. Huang, A.~Vaswani, J.~Uszkoreit, N.~Shazeer, C.~Hawthorne, A.~M. Dai, M.~D. Hoffman, and D.~Eck, ``An improved relative self-attention mechanism for transformer with application to music generation,'' \emph{arXiv preprint arXiv:1809.04281}, vol.~2, 2018.

\bibitem{liang2017automatic}
F.~T. Liang, M.~Gotham, M.~Johnson, and J.~Shotton, ``Automatic stylistic composition of bach chorales with deep lstm.'' in \emph{ISMIR}, 2017, pp. 449--456.

\end{thebibliography}


\section{Ethics Statement}
This particular work has no direct negative ethical implications. 

We acknowledge that Heinrich Schenker (the inventor of Schenkerian analysis) was racist and nationalist. His sociopolitical views are not condoned by authors of the present work. As originally designed, his style of analysis did not extend far beyond German composers of the common practice era. As we address in Section \ref{sec:introduction}, we do not see Schenkerian analysis as a static analysis defined by Schenker; rather, we see it as a growing and developing set of tools that may be applied to various musical genres, detached from Schenker's personal views. 

\end{document}